\documentclass[12pt,english,english,prb,preprint]{revtex4}
\usepackage{mathptmx}
\usepackage[T1]{fontenc}
\usepackage[latin9]{inputenc}
\usepackage{subfigure}
\usepackage{float}
\usepackage{amsmath}
\usepackage{graphicx}
\usepackage{amssymb}

\makeatletter

\usepackage{mathptmx}

\usepackage{subfigure}

\usepackage{babel}
\makeatother

\begin{document}

\title{A gradient-directed Monte Carlo approach to molecular design}

\author{Xiangqian Hu, David N. Beratan{*} and Weitao Yang{*}}

\affiliation{Department of Chemistry, Duke University, Durham, NC 27708-0354}

\date{\today}

\email{david.beratan@duke.edu}

\email{weitao.yang@duke.edu }

\begin{abstract}
The recently developed linear combination of atomic potentials (LCAP)
approach {[}M. Wang \textit{et al}., J. Am. Chem. Soc., \textbf{128},
3228 (2006)] allows continuous optimization in discrete chemical space
and thus is quite useful in the design of molecules for targeted properties.
To address further challenges arising from the rugged, continuous
property surfaces in the LCAP approach, we develop a gradient-directed
Monte Carlo (GDMC) strategy as an augmentation to the original LCAP
optimization method. The GDMC method retains the power of exploring
molecular space by utilizing local gradient information computed from
the LCAP approach to jump between discrete molecular structures. It
also allows random Monte Carlo moves to overcome barriers between
local optima on property surfaces. The combined GDMC and LCAP approach
is demonstrated here for optimizing nonlinear optical (NLO) properties
in a class of donor-acceptor substituted benzene and porphyrin frameworks.
Specifically, one molecule with four nitrogen atoms in the porphyrin
ring was found to have a larger first hyperpolarizability than structures
with the conventional porphyrin motif. 
\end{abstract}
\maketitle
\newpage{}

\section{Introduction}

Molecular and material design presents an extremely challenging problem
because the number of accessible stable candidate molecules in the
sampling space is immense\citep{Lipinski2004855-861,Ertl2003374-380,Dobson2004824-828}.
In such a vast molecular space, direct enumeration and evaluation
of molecules are expensive and inefficient. Furthermore, since each
molecular structure in the space is unique and discrete, traditional
methods are usually based on combinatorial chemistry. Each molecule
is generated from building blocks and its specific chemical properties
are analyzed for selection in the design process. Thus, a number of
discrete optimization approaches, such as branch-and-bound methods\citep{Ostrovsky2002645-660},
Monte Carlo, and genetic algorithms\citep{Wales19991368-1372,Kuntz1994117-123,Jorgensen20041813-1818},
have been utilized to explore the rich molecular space for the desired
chemical properties. All of these approaches require extensive molecular
property computations to obtain some optimal structures. To overcome
the difficulties associated with discrete optimization, several approaches\citep{Franceschetti199960-63,Kuhn199610595-10599}
have been proposed to transform the discrete optimization into a continuous
optimization. However, none of these methods can be generalized to
optimize the desired properties for molecular design. Recently, we
introduced a linear combination of atomic potentials (LCAP) scheme
\citep{Wang20063228-3232}. The LCAP method provides a smooth chemical
property surface interpolating those of the discrete target molecules,
thus facilitates property optimization for molecular design. A related
approach has also been suggested but was based on atom type variation
in a fixed chemical skeleton\citep{Rothlisberger2005153002}. 

A molecule or solid is characterized by its electron-nuclear potential

\begin{eqnarray}
v(\mathbf{r})=-\sum_{\mathbf{R}}\frac{Z_{\mathbf{R}}}{|\mathbf{r}-\mathbf{R}|}\label{lcap:vSch}\end{eqnarray}
where $Z_{\mathbf{R}}$ is the atomic number at position $\mathbf{R}$.
This electrostatic potential and the number of electrons $N$, enter
the Schrödinger equation and determine the properties of the molecules.
As such, molecular or material properties are functionals of $v(\mathbf{r})$
and $N$ . 

In the LCAP approach, we thus formulate the design of optimized molecules
or materials as the optimization of properties in the functional space
of $v(\mathbf{r})$ and $N$ \citep{Wang20063228-3232}. The advantages
of optimization based on the potential arise from both the potential\textquoteright{}s
\textquotedblleft{}smoothness\textquotedblright{} and the favorable
scaling of the computational cost with system size. The complexity
of the potential function grows linearly with the molecular size.
This is in stark contrast to the combinatorial explosion of possible
molecular structures that would fill a growing molecular volume. The
challenge at hand is how best to carry out the potential-function
optimization; it is essential that the optimized potential be linked
to real molecules. While all molecules lie within the space of all
$v(\mathbf{r})$, not all potentials map back to chemical structures.
A full optimization in potential space most likely will lead to a
potential that does not map to a molecule, which has a limited form
of the potential given by eq. (\ref{lcap:vSch}). 

The LCAP method is an effective way to limit the scope of the optimization
in potential space. Specifically, the potential is expanded as a linear
combination of atomic potentials, 

\begin{eqnarray}
v(\mathbf{r})=\sum_{\mathbf{R},A}b_{A}^{\mathbf{R}}v_{A}^{\mathbf{R}}(\mathbf{r})\label{lcap:v}\end{eqnarray}
where $v_{A}^{\mathbf{R}}(\mathbf{r})$ is the potential of an atom
(or chemical group) $A$ at position $\mathbf{R}$, $v_{A}^{\mathbf{R}}(\mathbf{r})=\sum_{B}v_{B}(\mathbf{r})$,
$v_{B}(\mathbf{r})=-\tfrac{Z_{\mathbf{R}}}{|\mathbf{r}-\mathbf{R}|}$.
The coefficient $b_{A}^{\mathbf{R}}$ defines the admixture of a particular
atom or group in the molecular potential. The constraints on $b_{A}^{\mathbf{R}}$
are $\sum_{A}b_{A}^{\mathbf{R}}=1$ and $0\leq b_{A}^{\mathbf{R}}\leq1$.
When $b_{A}^{\mathbf{R}}=0$ or $b_{A}^{\mathbf{R}}=1$ for each $\mathbf{R}$,
$v(\mathbf{r})$ corresponds to a {}``real'' molecule. Therefore,
within the LCAP framework, the design of molecules with an optimized
targeted property becomes the optimization of the coefficients $\{b_{A}^{\mathbf{R}}\}$
for given sets of $\mathbf{R}$ (geometry) and $v_{A}^{\mathbf{R}}$
(atom or group types). The LCAP scheme has been used to find the potential
function that produces structures with optimal properties, such as
electronic polarizabilities and first hyperpolarizabilities for the
simple case with two substitution sites. We showed that the optimal
structures can be determined without directly enumerating all of the
possible structures. We further implemented the LCAP scheme in the
tight-binding and semi-empirical frameworks\citep{Xiao2008,Keinan2007176-181}.
Our studies showed that the linear combination of atomic/group coefficients
in LCAP indeed smooths out the discreteness of the molecular space
and facilitates continuous optimization. 

However, when molecular diversity grows (e.g., as chemical groups
of varying structure, number of electrons, etc. enter), the ruggedness
of the property surfaces grows as well. The rugged surface makes the
property optimizations inefficient, since the optimization becomes
easily trapped in local minima. In addition, a fractional number of
electrons may appear during the continuous optimization process, which
can lead to convergence difficulty for the property calculations when
using quantum mechanical approaches. 

In this paper, we develop an optimization approach, the gradient-directed
Monte Carlo (GDMC) approach, which retains the convenience of continuous
optimization while circumventing some of the difficulties associated
with the ruggedness of property surfaces. The method utilizes local
gradient information with respect to the LCAP coefficients $\{b_{A}^{\mathbf{R}}\}$,
but preserves the simplicity of exploring molecular space by jumping
between discrete molecular structures. It also allows random Monte
Carlo jumps to overcome barriers between local property minima. As
such, the GDMC method avoids the surface ruggedness that arises from
chemical structures of {}``intermediate'' composition (i.e., structures
with non-integer LCAP coefficients; see Ref. \citep{Wang20063228-3232}
for further information). Note that the application GDMC in conjunction
with using finite difference approximation to the gradients was reported
in Ref. \citet{Keinan2007176-181}, without a description of algorithmic
details. The related LCAP application to locate the deepest minimum
energy configuration of face centered cubic Au - Pd alloys has also
been reported.\citep{2007402201-402207} In a related manner, the
trial wave function in a fixed-node quantum Monte Carlo method\citep{Reynolds19825593}
{}``serves as a guiding function for a random walk of the electrons
through configuration space'', which is indeed a biased MC approach
for the continuous electron coordinate space. This is similar with
our proposed GDMC approach for the discrete molecular space.

We apply the combined GDMC and LCAP approaches to optimize the static
first electronic hyperpolarizability $\beta$ for molecular libraries.
In Section \ref{sec:LCAP}, the LCAP approach for the first hyperpolarizability
is explained briefly. In Section \ref{sec:GDMC}, the general GDMC
algorithm is introduced for molecular property optimization. Section
\ref{sec:LCAPGDMC} describes how to combine GDMC with the LCAP method
to optimize the first electronic hyperpolarizability. Three examples
of hyperpolarizability optimization are presented in Section \ref{sec:results}.
Finally, Section \ref{sec:Conclusion} summarizes the combined GDMC
and LCAP approaches.

\section{The LCAP approach for the first hyperpolarizability \label{sec:LCAP}}

Traditional molecular design is carried out by examining families
of discrete structures, which may be generated using combinatorial
libraries, intuition, or qualitative structure-activity relationships.
But for any molecule, the Schrödinger equation is determined by the
nuclear-electron attraction potential $v(\mathbf{r})$ and the number
of electrons $N$. Optimization of $v(\mathbf{r})$, as developed
in our recent studies\citep{Wang20063228-3232}, is appealing because
continuous search algorithms based on gradient calculations can be
much more efficient than traditional molecular design methods. The
discrete molecular optimization problem becomes a continuous one when
$v(\mathbf{r})$ is expanded as a linear combination of atomic potentials
(LCAP), as in Eq.( \ref{lcap:v}).

In this paper, we are interested in the linear and nonlinear response
properties to an external electric field. In an electric field $\mathbf{F}$,
the molecular energy is \begin{eqnarray}
E(\mathbf{F}) & = & E(0)-\sum_{i}\mu_{i}F_{i}-\frac{1}{2!}\sum_{ij}\alpha_{ij}F_{i}F_{j}\nonumber \\
 & - & \frac{1}{3!}\sum_{ijk}\beta_{ijk}F_{i}F_{j}F_{k}-\frac{1}{4!}\sum_{ijkl}\gamma_{ijkl}F_{i}F_{j}F_{k}F_{l}-\cdots\label{energyfield}\end{eqnarray}
 where $E(0)$ is the energy in the absence of the applied electric
field, the indices $i$, $j$, $k$, and $l$ span the molecular ($x$,
$y$, and $z$) axes, $\mu_{i}$ is the permanent dipole moment and
$\alpha_{ij}$ is the (linear) polarizability. The first hyperpolarizability
is $\beta_{ijk}$ and the second hyperpolarizability is $\gamma_{ijkl}$.
We focus on the static first hyperpolarizability in this paper. Specifically,
the first hyperpolarizability tensor element $\beta_{zzz}$ along
the z-axis can be calculated using the finite-field approach \citep{Kurtz1998241,Kurtz199082-87}:
\begin{eqnarray}
\beta_{zzz}=\left[-\frac{1}{2}\left\{ E(2F_{z})-E(-2F_{z})\right\} +\left\{ E(F_{z})-E(-F_{z})\right\} \right]/F_{z}^{3}\label{lcap:bzzz}\end{eqnarray}
where $F_{z}$ is the applied electric field along the z-axis and
$E(F_{z})$ is the corresponding energy.

In the LCAP approach, the $\beta_{zzz}$ gradients with respect to
the LCAP coefficients $\{b_{A}^{\mathbf{R}}\}$ are,

\begin{eqnarray}
\frac{\partial\beta_{zzz}}{\partial b_{A}^{\mathbf{R}}} & = & \left\{ -\frac{1}{2}\left[\frac{\partial E(2F_{z})}{\partial b_{A}^{\mathbf{R}}}-\frac{\partial E(-2F_{z})}{\partial b_{A}^{\mathbf{R}}}\right]+\left[\frac{\partial E(F_{z})}{\partial b_{A}^{\mathbf{R}}}-\frac{\partial E(-F_{z})}{\partial b_{A}^{\mathbf{R}}}\right]\right\} /F_{z}^{3}\nonumber \\
 & = & \left\{ -\frac{1}{2}\left[\left.\frac{\partial E(2F_{z})}{\partial b_{A}^{\mathbf{R}}}\right|_{N}-\left.\frac{\partial E(-2F_{z})}{\partial b_{A}^{\mathbf{R}}}\right|_{N}\right]+\left[\left.\frac{\partial E(F_{z})}{\partial b_{A}^{\mathbf{R}}}\right|_{N}-\left.\frac{\partial E(-F_{z})}{\partial b_{A}^{\mathbf{R}}}\right|_{N}\right]\right\} /F_{z}^{3}+\nonumber \\
 &  & \left\{ -\frac{1}{2}\left[\left.\frac{\partial E(2F_{z})}{\partial b_{A}^{\mathbf{R}}}\right|_{v}-\left.\frac{\partial E(-2F_{z})}{\partial b_{A}^{\mathbf{R}}}\right|_{v}\right]+\left[\left.\frac{\partial E(F_{z})}{\partial b_{A}^{\mathbf{R}}}\right|_{v}-\left.\frac{\partial E(-F_{z})}{\partial b_{A}^{\mathbf{R}}}\right|_{v}\right]\right\} /F_{z}^{3}\label{lcap:bgrad}\end{eqnarray}
as implemented\citep{Wang20063228-3232} in the PWSCF\citep{PWSCF}
program based on plane waves with ultra-soft pseudopotentials (USPP)\citep{Vanderbilt19907892-7895}.
In PWSCF, for atom $A$, $v_{A}^{\mathbf{R}}(\mathbf{r})$ is replaced
by the USPP, \begin{eqnarray}
v_{A}(\mathbf{r},\mathbf{r}') & = & v_{A}^{loc}(\mathbf{r},\mathbf{r}')+v_{A}^{nl}(\mathbf{r},\mathbf{r}')\nonumber \\
 & = & v_{A}^{loc}(r)\delta(\mathbf{r}-\mathbf{r}')+\sum_{nm}D_{nm}^{0}|\beta_{n}^{A}\left>\right<\beta_{m}^{A}|\label{lcap:vauspp}\end{eqnarray}
where $D_{nm}^{0}$, $\beta_{n}^{A}$ and $v_{A}^{loc}(r)$ characterize
the USPP for atom $A$. The electrostatic potential $v(\mathbf{r})$
can be expanded as a linear combination of atomic potentials in the
USPP framework, \begin{eqnarray}
v(\mathbf{r},\mathbf{r}') & = & \sum_{\mathbf{R},A}b_{A}^{\mathbf{R}}v_{A}^{\mathbf{R},loc}(r)\delta(\mathbf{r}-\mathbf{r}')+\sum_{\mathbf{R},A}b_{A}^{\mathbf{R}}v_{A}^{\mathbf{R},nl}(\mathbf{r},\mathbf{r}')\nonumber \\
 & = & v^{loc}(\mathbf{r},\mathbf{r}')+v^{nl}(\mathbf{r},\mathbf{r}')\label{lcap:vuspp}\end{eqnarray}
where the first term is the total local potential and the second term
is the total nonlocal potential of $v(\mathbf{r})$ for the system.
The total energy of the system in an external electric field can then
be written \begin{eqnarray}
E(F_{z}) & = & \sum_{i}\left<\phi_{i}\left|-\frac{\hbar}{2m}\nabla^{2}+v^{nl}(\mathbf{r},\mathbf{r}')\right|\phi_{i}\right>+E_{H}\left[n(\mathbf{r})\right]\nonumber \\
 & + & E_{xc}\left[n(\mathbf{r})\right]+\int d\mathbf{r}\left(v^{loc}(\mathbf{r})-zF_{z}\right)n({\mathbf{r}})\label{lcap:energy}\end{eqnarray}
where \begin{eqnarray}
n(\mathbf{r}) & = & \sum_{i}\left[|\phi_{i}|^{2}+\sum_{\mathbf{R},A}b_{A}^{\mathbf{R}}\sum_{nm}Q_{nm}^{A}\left<\phi_{i}|\beta_{n}^{A}\right>\left<\beta_{m}^{A}|\phi_{i}\right>\right]\label{lcap:rho}\\
\left<\phi_{i}|S|\phi_{j}\right> & = & \delta_{ij},S=1+\sum_{\mathbf{R},A}\sum_{nm}b_{A}^{\mathbf{R}}q_{nm}^{A}|\beta_{n}^{A}\left>\right<\beta_{m}^{A}|.\label{lcap:overlap}\end{eqnarray}
Here, $q_{nm}^{A}=\int d\mathbf{r}Q_{nm}^{A}(\mathbf{r})$. From eqs.
\ref{lcap:energy}, \ref{lcap:rho}, \ref{lcap:overlap}, the derivative
of the total energy with respect to the coefficients $b_{A}^{\mathbf{R}}$
with $N$ electrons is: \begin{eqnarray}
\frac{\partial E(F)}{\partial b_{A}^{\mathbf{R}}}=\left.\frac{\partial E(F)}{\partial b_{A}^{\mathbf{R}}}\right|_{N}+\left.\frac{\partial E(F)}{\partial b_{A}^{\mathbf{R}}}\right|_{v(\mathbf{r},\mathbf{r}')}\label{lcap:grad}\end{eqnarray}
 where \begin{eqnarray}
\left.\frac{\partial E(F)}{\partial b_{A}^{\mathbf{R}}}\right|_{N} & = & \sum_{i}\sum_{nm}D_{nm}^{A}\left<\phi_{i}|\beta_{n}^{A}\right>\left<\beta_{m}^{A}|\phi_{i}\right>+\int v_{A}^{\mathbf{R},loc}(\mathbf{r})n(\mathbf{r})d\mathbf{r}\nonumber \\
 & - & \sum_{i}\epsilon_{i}\sum_{nm}q_{nm}^{A}\left<\phi_{i}|\beta_{n}^{A}\right>\left<\beta_{m}^{A}|\phi_{i}\right>\label{lcap:grad1}\\
\left.\frac{\partial E(F)}{\partial b_{A}^{\mathbf{R}}}\right|_{v=v(\mathbf{r},\mathbf{r}')} & = & \frac{\partial E(F)}{\partial N}\frac{\partial N}{\partial b_{A}^{\mathbf{R}}}=\mu Z_{A}^{\mathbf{R}}\label{lcap:grad2}\end{eqnarray}
Here, $D_{nm}^{A}=D_{nm}^{0}+\int V_{eff}(\mathbf{r})Q_{nm}^{A}(\mathbf{r})d\mathbf{r}$,
$V_{eff}(\mathbf{r})=V_{H}(\mathbf{r})+V_{xc}(\mathbf{r})+v^{loc}(\mathbf{r})-zF$,
and $\sum_{\mathbf{R},A}Z_{A}^{\mathbf{R}}=N$. The chemical potential
of the system is $\mu$ , and $Z_{A}^{\mathbf{R}}$ is the atomic
number of atom $A$, or the electron number of functional group $A$
at position $\mathbf{R}$. After obtaining the gradients $\{\frac{\partial E}{\partial b_{A}^{\mathbf{R}}}\}$,
the $\beta_{zzz}$ gradients with respect to LCAP coefficients can
be computed using eq. \ref{lcap:bgrad}. As described in Ref. \citep{Wang20063228-3232},
with the availability of gradients, continuous optimization can be
carried out effectively.

Our studies with expanded molecular fragment libraries showed that
the property surfaces can become too complicated to permit efficient
hyperpolarizability optimization in the continuous space. Indeed,
for the specific NLO properties, potentials associated with non-integer
LCAP coefficients can have large and rapidly varying hyperpolarizabilities
as electron numbers are varied and as electronic states move in and
out of degeneracy. This challenge presented by adding richness to
the LCAP library suggests that we could modify the optimization method
in a way that retains the power and appeal of the LCAP approach while
side-stepping the ruggedness associated with the non-integer (not
realizable, or {}``intermediate'') molecular structures. In the
following section, we present the development in which the GDMC and
LCAP methods are combined to overcome the difficulties associated
with the ruggedness of the property surface.

\section{The general GDMC approach\label{sec:GDMC}}

For a given discrete space, the discrete optimization problem may
be transformed into be a continuous one by interpolation \citep{Wang20063228-3232,Koh2005728,Koh2005109-130}.
Then, the interpolated continuous surface can be used for the specific
optimization problem of interest. For example, in molecular design,
when one fragment $A$ is replaced by another fragment $B$ to form
a new molecule, this transformation is indeed discrete. Additionally,
the weighting of each fragment at each site is defined by one coefficient.
When the molecule contains $A$, the coefficient of $A$ is set to
1; otherwise it is set to 0. The set of coefficients (either 0 or
1) corresponds to one real molecule. Thus, normal integer programming
algorithms such as Monte Carlo and genetic algorithms can be applied
to search in the discrete space. However, we can also allow $A$ to
change continuously to $B$, which means the $A$ coefficient is decreased
from 1 to 0 continuously and the $B$ coefficient is increased from
1 to 0. It is then possible to use gradients with respect to the coefficients
to aid the molecular property optimization. Mathematically, the LCAP
coefficients discussed in Section \ref{sec:LCAP} play the same roles
in the potential optimization to transform a discrete problem into
a continuous one. The possible transformations are not unique. Some
approaches may simplify the optimization problem while others may
worsen the problem, depending on the property and the specific molecular
fragment libraries. In the LCAP approach, when the coefficients are
between 0 and 1, the system becomes an {}``intermediate'' or alchemical
species. It may lead to non-smooth behavior of the molecular property
in the unphysical intermediate states. Such a scenario was encountered
in the LCAP continuous optimization due to the fractional number of
electrons when the NLO properties were targeted for optimization.
We have found that the system with one fractional electron is more
likely to have a much higher nonlinear polarizability than the corresponding
systems with an integer number of electrons because the NLO property
is quite sensitive to the electronic states. Thus, it may be particularly
ineffective to optimize NLO properties for diverse libraries on an
interpolated continuous surface. 

Although the interpolated continuous surface may be rugged, the local
gradient information for each set of coefficients represents approximately
how the property varies when the coefficients change. For example,
as demonstrated in our previous studies\citep{Keinan2007176-181},
calculated gradients for the real molecule were used to build the
next molecule and to guide the search efficiently. But the search
can be easily trapped in local optima. Therefore, we develop a new
strategy, the gradient-directed Monte Carlo (GDMC) approach, to deal
with the global optimization of discrete variables. GDMC utilizes
the local gradients to enhance the search for local optima, while
the stochastic jumps overcome the barriers between local optima. The
details of a GDMC calculation for the minimization of a general property
$P$ consist of the following steps :

\begin{enumerate}
\item Set the iteration number $i=0$ and begin with an initial set of coefficients
$\{b_{A}^{1}\}$ (either 0 or 1) at each site. 
\item Set $i=i+1$. Calculate the property $P_{i}$ and its gradients $\frac{\partial P_{i}}{\partial b_{A}^{i}}$
with respect to all coefficients $\{b_{A}^{i}\}$ at each site. Exit
if the optimization goal or the maximum number of iteration is reached.
\item Make a trial move to generate a new set of $\{b_{A}^{i}\}$ following
the computed gradients $\{\frac{\partial P_{i}}{\partial b_{A}^{i}}\}$.
Accept the trial move with the probability $p=min\{1,\: e^{-\beta(P_{i}-P_{i-1})}\}$
using the Metropolis rule. If the trial move is accepted, go to step
2; If the trial move is rejected, go to step 3 . 
\end{enumerate}
Here, $\beta=1/(kT)$. On the energy surface, $k$ is Boltzmann's
constant and $T$ is the temperature. However, on the property surface,
$k$ and $T$ are two parameters: the value of $k$ is constant; $T$
is an empirical temperature that controls the probability of accepting
a set of coefficients. The rules for generating new trial molecules
following the LCAP gradients are described in the next section. 

Compared to the classical MC method, GDMC does not randomly vary the
coefficients. All of the coefficients (either 0 or 1) that describe
the discrete space are generated from the property gradients that
can be defined in many ways. As shown in Section \ref{sec:LCAPGDMC},
we redefined the LCAP gradients associated with GDMC to optimize the
NLO properties in three different examples. It is also worth noting
that the GDMC approach is suitable for any discrete optimization that
possesses a smoothed-out virtual continuous surface.

\section{Combining GDMC and LCAP\label{sec:LCAPGDMC}}

We use the absolute value of the first hyperpolarizability as the
object function in the GDMC optimization. For the LCAP approach, the
gradients $\{\frac{\partial\beta_{zzz}}{\partial b_{A}^{\mathbf{R}}}\}$
(eq. \ref{lcap:bgrad}) include the chemical potential $\mu=\left(\frac{\partial E}{\partial N}\right)_{v}$
(see eq. \ref{lcap:grad2}), which has two values for each of the
directions of change in the number of electrons and can be calculated
based on the formula recently derived for any DFT calculations.\citep{Cohen2008115123-115128}
Here, we found it possible to bypass the need to calculate $\mu$.
We redefine the LCAP gradients by a normalization, \begin{eqnarray}
\frac{\partial\beta_{zzz}}{\partial b_{A}^{\mathbf{R}}}' & = & \frac{\partial\beta_{zzz}}{\partial b_{A}^{\mathbf{R}}}/Z_{A}^{\mathbf{R}}\nonumber \\
 & = & (\left.\frac{\partial\beta_{zzz}}{\partial b_{A}^{\mathbf{R}}}\right|_{N})/Z_{A}^{\mathbf{R}}\nonumber \\
 & + & \left\{ -\frac{1}{2}\left[\mu_{chem}(2F_{z})-\mu_{chem}(-2F_{z})\right]+\left[\mu_{chem}(F_{z})-\mu_{chem}(-F_{z})\right]\right\} /F_{z}^{3}.\label{gdmc:grad}\end{eqnarray}
 where $Z_{A}^{\mathbf{R}}$ is the total number of electrons for
atom or functional group $A$ at position $\mathbf{R}$. Thus, all
gradients are simply shifted by a constant value, defined in the last
term of eq. \ref{gdmc:grad}, that comes from the chemical potential.
Because this term does not change the ordering of the gradient values,
it is set to zero. During the GDMC optimization, one molecule is generated
at a time and the gradients $\{\frac{\partial\beta_{zzz}}{\partial b_{A}^{\mathbf{R}}}'\}$
are calculated for that structure using eq. \ref{lcap:grad}, \ref{lcap:grad1},
\ref{lcap:grad2}, \ref{lcap:bgrad} and \ref{gdmc:grad}. These gradients
are then used to generate the next molecule. 

How the calculated gradients are used to obtain the next molecule
within the LCAP framework is critical. The main idea is to use the
LCAP gradients to choose the most promising fragment at each site.
This scheme is best explained with a simple example for a framework
with two fragment sites shown in Fig. \ref{lcapgdmc:rules}. Each
site has a library of four possible fragments. The current molecule
is $A1B1$ as seen in Fig. \ref{lcapgdmc:rules}. The LCAP gradients
$\{\frac{\partial\beta_{zzz}}{\partial b_{A}^{\mathbf{R}}}'\}$ are
calculated for each fragment. For each site, all four functional groups
are sorted in descending order according to their negative gradient
values ($\{-\frac{\partial\beta_{zzz}}{\partial b_{A}^{\mathbf{R}}}'\}$).
In Fig. \ref{lcapgdmc:rules}, the order after sorting becomes $A3>A1>A2>A4$
for site $A$, and $B2>B4>B1>B3$ for site $B$. Based on the ordering
at each site, the next molecule is $A3B2$. If $A3B2$ has been visited
before, one of the sites is chosen at random to undergo further mutation.
As shown in Fig. \ref{lcapgdmc:rules}, fragment $A3$ at site $A$
is replaced by fragment $A1$ (the next highest ranking fragment)
and the new molecule $A1B2$ is generated. This procedure is repeated
until a new molecule is obtained. However, because \textit{ab-initio}
calculations of the first hyperpolarizabilities are expensive, each
design site will be mutated only once. As such, if a new molecule
cannot be generated after several trial mutations or if the maximum
number of iterations has been reached, the optimization stops. Using
this procedure to generate a new molecule according to the normalized
LCAP gradients, it is straightforward to implement the GDMC algorithm
described in Section \ref{sec:GDMC} to optimize NLO properties. The
new GDMC-LCAP approach is tested in Section \ref{sec:results}. 

In molecular design, the geometry changes of the target system have
important effects on the target properties. In the current work, since
we focus on optimizing the first hyperpolarizabilities of the extended
$\pi$-electron structures with donors and acceptors, all of the candidate
molecular structures are rigid. The substitutions at design sites
do not disrupt the $\pi$-conjugation. Therefore, during the GDMC
optimizations for three cases in \ref{sec:results}, the framework
geometries were fixed.

\section{GDMC-LCAP optimization of the First hyperpolarizability \label{sec:results}}

Extended $\pi$-electron structures with donors and acceptors are
well known to have large first hyperpolarizabilities\citep{Chen19943117-3118,Cheng199110631-10643,Cheng199110643-10652,Dehu19936198-6206,Di1993682-686,Garito19811-26,Kanis199410089-10102,Kanis1994195-242,Lalama1981940-942,Li19881707-1715,Marder19933006-3007,Meyers199410703-10714,Stiegman19917658-7666,Suslick19926928-6930,Misra2005108-117},
and their response properties are well understood in the context of
both simple models and more involved quantum chemical calculations
\citep{Kanis1994195-242}. Thus, our studies focus on donor-$\pi$-acceptor
frameworks and one dominant tensor element, $\beta_{zzz}$, is computed
and optimized. To explore the efficiency of the GDMC-LCAP approach,
we optimized the absolute value of the first hyperpolarizability for
three different $\pi$-electron conjugated scaffolds. The size of
the cubic box in PWSCF\citep{PWSCF} was $22\times22\times22\ au^{3}$
in Case I, $30\times30\times30\ au^{3}$ in Case II, and $60\times60\times60\ au^{3}$
in Case III. The energy cutoff in the PWSCF calculation was 15 Ryd.\citep{cutoff}
The step size of the electric fields in Case I was $0.005au$ and
was set to $0.0025au$ in Cases II and III (the smaller step size
was used because of convergence issues during the SCF calculations).
The functional used was LDA\citep{Perdew19815048-5079} and the maximum
number of iterations was 30 except Case III, where it was 40. The
empirical temperature parameter was also studied in three cases.

\subsection{Case I}

We first used a six-membered ring system to test the GDMC-LCAP optimization
approach. The structural framework is shown in Fig.\ref{case1:frame}.
The geometry was taken from benzene ($r_{C-C}=1.390\AA$, $r_{C-H}=1.095\AA$)
and was held fixed during the optimization. Either a $CH$ group or
$N$ atom could be placed at each site, producing 64 possible structures
(many are identical due to the rotational symmetry). The structures
can also be enumerated to determine the largest $\beta_{zzz}$ value.
From the enumeration, structure (b) in Fig. \ref{case1:opt} was found
to have the largest $\beta_{zzz}$ value.

The GDMC-LCAP optimization began with the initial molecule, structure
(a) in Fig. \ref{case1:opt}, and the optimization profile is shown
in Fig. \ref{case1:opt}. Five runs with different initial molecules
were tested as well. They all yielded similar search profiles and
produced structure (b). In Fig. \ref{case1:opt}, three degenerate
molecules that had the same structure (b) (due to symmetry) were found.
Since the initial structure (a) has a relatively large $\beta_{zzz}$
value, the optimization becomes trapped in this local optimum and
the Monte Carlo stochastic moves were helpful in generating the next
molecule and overcoming the barrier between local optima. From the
$6th$ molecule in the profile, the LCAP gradients guided the molecular
generation quite efficiently. In molecules $7-10$ generated from
the LCAP gradient information, the absolute value of $\beta_{zzz}$
increased until the maximum value of $\beta_{zzz}$ was reached. The
same behavior, led by local optimization based on the LCAP gradients,
is also observed in the later steps shown in Fig. \ref{case1:opt}.
The total number of molecular property calculations was only 26, and
14 jumps between molecules based on the LCAP gradients showed increased
$\beta_{zzz}$ values. This indicates that the LCAP gradients indeed
helped generating new molecules with increased property values.

\subsection{Case II}

In Case II, a more complicated family of structures was explored.
The structural framework is shown in Fig. \ref{case2:frame}. The
six-member ring geometry is the same as in Case I. The geometries
of the functional groups were generated using the Spartan Builder
\citep{spartan}. Either an $H$, $CH_{3}$, $NO_{2}$ or $NH_{2}$
fragment was placed at sites 1 and 2. Either a $CH$ or $N$ was placed
at sites 3 through 6. Sites 1 and 2 are located on the z-axis. By
enumerating all 256 possible structures, molecule (b) was found to
have the largest $\beta_{zzz}$ value, shown in Fig. \ref{case2:opt}. 

In Case II, at sites 1 and 2, the four functional groups have different
numbers of electrons. When the continuous LCAP optimization is implemented,
we observed that some {}``intermediate'' molecules with a fractional
number of electrons had large $\beta_{zzz}$ values. The fractional
number of electrons is unphysical and makes the continuous surface
rugged. However, in the GDMC-LCAP optimization, only the properties
of real molecules are computed and the LCAP gradients direct the jumping
from one molecule to another. Since this surface is more complicated,
we tested the optimization using three different temperature parameters:
$T=1K$, $50K$, and $100K$. All the optimizations began with the
same initial molecule (a), shown in Fig. \ref{case2:opt}, and the
optimization profiles are shown in Fig. \ref{case2:opt}. 

All the GDMC-LCAP optimizations found the same molecule (b). When
$T=1K$, only 17 molecules were computed and the best molecule was
found in 2 jumps, indicated by the solid black line of Fig. \ref{case2:opt}
using the ordering of the LCAP gradients defined in Section \ref{sec:LCAPGDMC}.
After the properties of 17 molecules were calculated, the optimization
became trapped because of the low temperature. When the temperature
was increased from $1K$ to $50K$, the GDMC protocol quickly jumped
out of the local optimum around the initial molecule (a) in the beginning
of the optimization, as shown by the dotted red line in Fig. \ref{case2:opt}.
The largest $\beta_{zzz}$ molecule was obtained in the third property
calculation. As the optimization continued at the higher temperature,
GDMC generated new molecules that were more likely to be accepted
based on their property values. The other degenerate molecule (b)
was obtained in the eleventh property calculation. When the temperature
was increased further to $100K$, the profile showed similar behavior.
Basically, when $T$ varies from $1K$ to $100K$, GDMC has more flexibility
to jump out of local optima and explore the discrete space more broadly.

\subsection{Case III}

In Case III, a large porphyrin substituted donor-$\pi$-acceptor framework
was used (see Fig. \ref{case3:frame}). The geometry was taken from
molecule (c) in Fig. \ref{case3} that had been optimized at the $B3LYP/6-31+G*$
level using $Gaussian03$.\citep{g03} The geometries of all donors
and acceptors were generated using the Spartan Builder \citep{spartan}.
At site $A$, one of three acceptor groups ($NO_{2}$, $CN$ or $COH$)
was placed. At site $D$, one of three donor groups ($N(CH_{3})_{2}$,
$CH_{3}$ or $OCH_{3}$) was placed. Sites 1 through 10 have either
a $CH$ or $N$ unit. This family of structures includes a total of
9,216 possible molecules. Performing ab initio first hyperpolarizability
calculations on all of the possible structures is costly. Thus, we
cannot enumerate all of the structures to find the one with the largest
$\beta_{zzz}$ value.

The initial molecule (a) consists of a normal porphyrin ring with
a $CH_{3}$ at site $D$ and a $CN$ at site $A$ that were chosen
randomly from four possible donors and four acceptors. Based on $\beta_{zzz}$
calculations for several initial molecules, the $\beta_{zzz}$ values
vary from thousands to tens of thousands (in atomic units). Thus,
we executed three GDMC optimizations at three different temperatures
beginning with the initial molecule (a) shown in Fig. \ref{case3}.
The maximum number of iterations was 40. All of the optimization profiles
are shown in Fig. \ref{case3:opt}. Two of the trajectories were trapped
in local optima because the surface is rugged. The optimizations at
$T=300K$ and $T=500K$ were stuck after less than 25 molecules were
analyzed. At $T=800K$, GDMC jumped out of the local optimum and found
one better molecule (b), shown in Fig. \ref{case3}. However, we further
tested five trial runs beginning from five different random initial
molecules at $T=300K$. The most favorable molecule of all trial runs
was molecule (b), shown in Fig. \ref{case3}. This suggests that for
much more rugged surface, the optimization is easily trapped when
$T$ is not sufficiently high. Hence, for the low temperature such
as $T=300K$ in this case, several independent optimizations with
different random initial molecules allow more broad explorations of
the molecular space.  In addition, even for this rugged surface, the
local gradient information from LCAP assists in jumping from one molecule
to another with a higher $\beta_{zzz}$ value. For example, at $T=800K$,
the $\beta_{zzz}$ value was increased from molecules 21 to 25 in
the optimization profile.

 The obtained molecule (b) shown in Fig. \ref{case3} contains a new
type of porphyrin-like ring. The origin of its high $\beta_{zzz}$
becomes an interesting question. We speculate that, because nitrogen
has more electrons than carbon, the four additional nitrogen atoms
present in molecule (b) increase the number of $\pi$ electrons that
are delocalized in the ring. In addition, the asymmetry of the broken
ring facilitates polarization. Furthermore, molecule (b) has the strongest
donor ($N(CH_{3})_{2}$) and acceptor ($NO_{2}$)\citep{Kanis1994195-242}
among the simple donor-acceptor structures. For these reasons, we
predict that molecule (b) of Fig. \ref{case3} has a larger $\beta_{zzz}$
value than those with conventional porphyrin cores. Molecule (b) may
represent a global optimum or a near optimal structure.

\section{Conclusion\label{sec:Conclusion}}

Although the original LCAP approach was demonstrated to optimize the
first hyperpolarizability of simple systems efficiently\citep{Wang20063228-3232},
it suffers from the rugged surfaces when LCAP is applied for more
complicated molecular libraries. During the property optimization,
some {}``intermediate'' or {}``alchemical'' molecules were obtained
with unphysically large first hyperpolarizabilities because of an
unphysical fractional number of electrons. However, the LCAP gradients
of the property with respect to the LCAP coefficients are useful in
directing the search for candidate molecules. To utilize these LCAP
gradients and to avoid analyzing {}``intermediate'' molecules, we
developed a gradient-directed Monte Carlo (GDMC) method that can be
used in discrete optimization. After making the discrete space continuous,
the property gradients with respect to the discrete variables can
be calculated. The gradients are then used to generate a new set of
discrete variables. The stochastic Monte Carlo moves assist in jumping
out of local optima. Specifically, the LCAP gradients computed from
one real molecule are normalized and utilized to generate a new structure.
Therefore, the combined GDMC-LCAP approach was used to optimize the
first hyperpolarizability for three different structural frameworks. 

\begin{figure}[H]
\begin{centering}
\includegraphics[width=15cm]{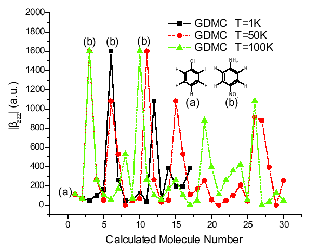} 
\par\end{centering}

\caption{}

\label{case2:opt} 
\end{figure}

\begin{figure}[H]
\begin{centering}
\subfigure[]{\includegraphics[width=5cm]{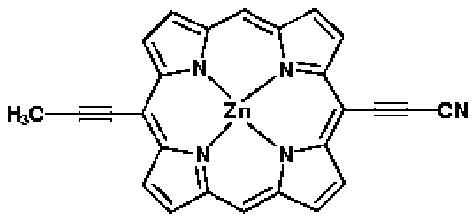}}\\
\subfigure[]{\includegraphics[width=5cm]{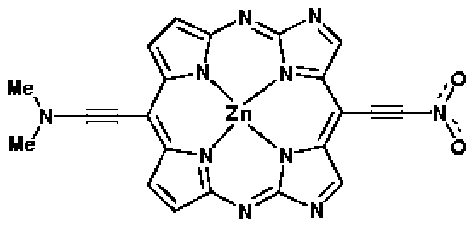}}\\
\subfigure[]{\includegraphics[width=5cm]{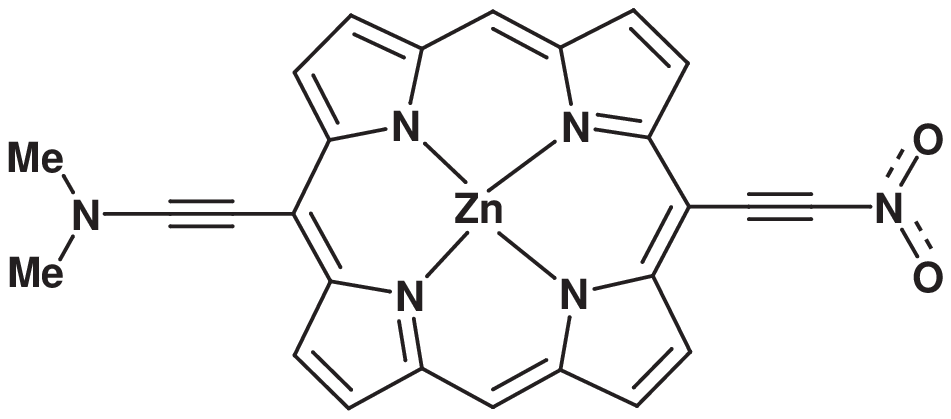}} 
\par\end{centering}

\caption{\label{case3}}

\end{figure}

For Cases I and II, several runs of the GDMC-LCAP approach always
found the most favorable structure that was validated by the exhaustive
enumeration of all possible molecules. The jumps from one molecule
to another based on the LCAP gradients have greater likelihood to
increase the molecular first hyperpolarizability.   The temperature
used in GDMC-LCAP is an empirical parameter. When $T$ is sufficiently
high, the optimization stops only if the maximum number of iterations
is reached, for example, see Fig. \ref{case2:opt} for Case II when
$T=50K$ and $T=100K$. When $T$ is too low, the optimization is
trapped after several trial mutations, for example, see Fig. \ref{case2:opt}
for Case II when $T=1K$ . To explore the molecular space more broadly
when $T$ is low and ensure that the optimal property is found, several
optimizations beginning from random initial molecules may be required.
For example, in Case III, five trial runs beginning from five different
random initial molecules at $T=300K$ were performed and structure
(b) in Fig. \ref{case3} was obtained in all five runs. Furthermore,
if GDMC is combined with heating, cooling, or annealing protocols,
GDMC will be more robust and irrespective of the initial guess.

In Case III,  the $\beta_{zzz}$ value of structure (b) in Fig. \ref{case3}
obtained during the GDMC optimizations is increased almost $20\%$
compared to the initial seed molecule (a) in Fig. \ref{case3}. In
particular, the optimized structure has the strongest donor and acceptor
groups at the $D$ and $A$ sites and a $\pi$-electron ring with
four nitrogen atoms more than in a porphyrin. Thus, we predict that
structure (b) in Fig. \ref{case3} is the optimal or nearly optimal
structure.

To compare the GDMC method with other approaches such as the classical
Monte Carlo (MC) method and a genetical algorithm (GA), we recently
studied an HP model\citep{Dill893986,Dill913775,Dill95561} for protein
sequence design and protein folding.\citep{Hu2008} Those results
indicate that the GDMC approach is much more efficient than the classical
MC or GA optimization, because the property gradients with respect
to the discrete variables, combined with the Monte Carlo moves, assist
in jumping out of local optima and produce a greater likelihood of
generating a new set of discrete variables with an enhanced property
value. 

When the geometry changes of frameworks are involved during the property
optimization, the advantage of local searching directed by property
gradients may not be clear. However, in a recent work, GDMC was employed
to optimize protein sequence with concurrent protein structure optimization.\citep{Hu2008a}
It shows that GDMC is still more efficient than the classical MC. 

In summary, the combined GDMC-LCAP approach provides an effective
strategy for finding molecules with optimized molecular properties.
The GDMC method can be used to explore other discrete spaces as well
as the LCAP space. 

\begin{acknowledgments}
Support from the DARPA Predicting Real Optimized Materials project
through the Army Research Office (ARO) is gratefully acknowledged
(W911NF-04-1-0243). DNB thanks the Keck and NEDO Foundations for support
of computational infrastructure. WY acknowledges partial support from
the National Science Foundation. X. H. thanks Jerry M. Parks and Hao
Hu for helpful discussions. 
\end{acknowledgments}
\bibliography{612830JCP}

\begin{thebibliography}{46}
\bibitem[{cut()}]{cutoff}
\bibinfo{note}{Three different energy cut off (i.e., 15, 25, and 35 Ryd) in
  PWSCF were tested for several molecules using ultra-soft pseudopotential and
  15 Ryd was appropriate to compute the first hyperpolarizability.}
\end{thebibliography}

\begin{figure}[H]
\begin{centering}
\includegraphics[width=10cm]{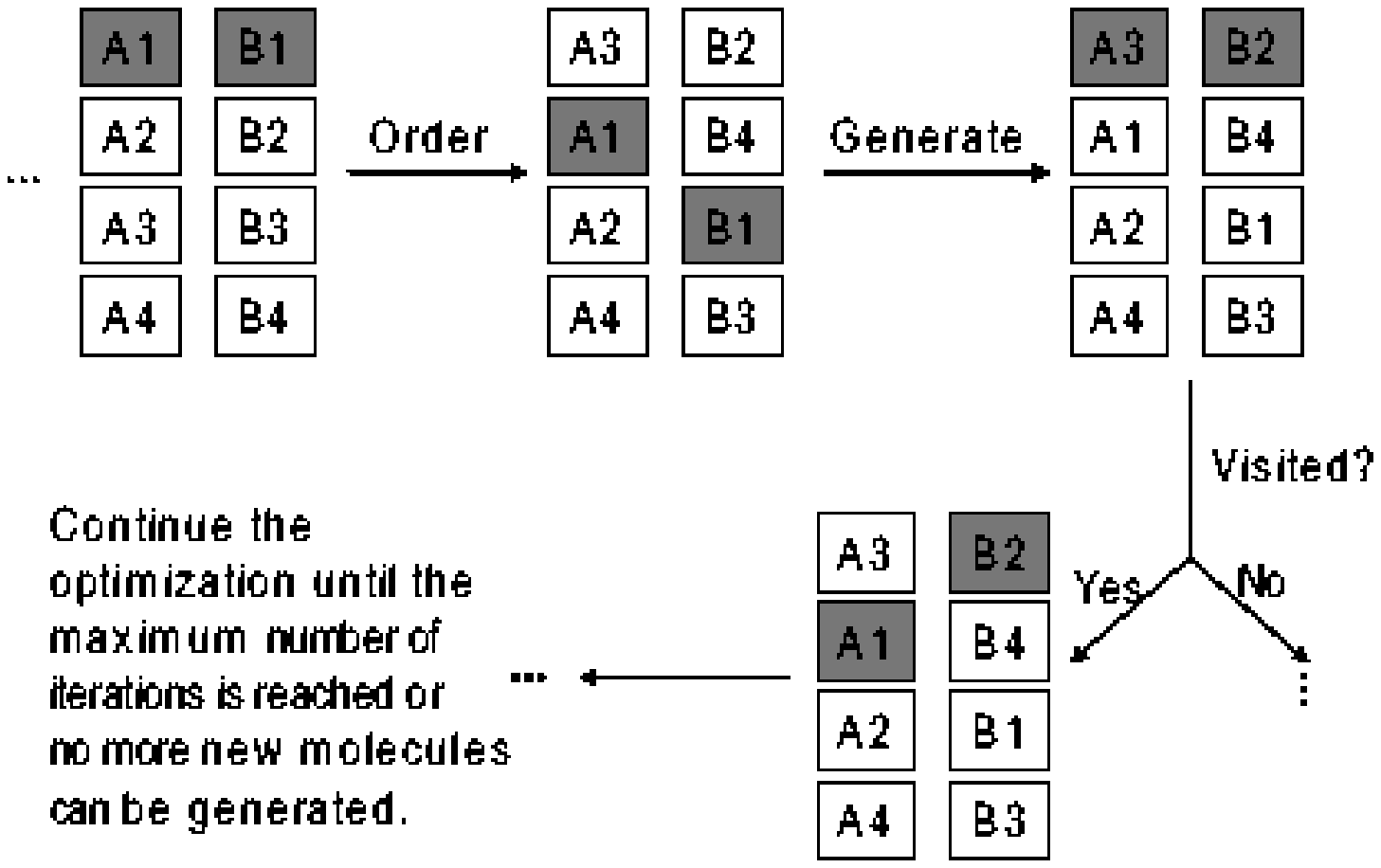} 
\par\end{centering}

\caption{}

\label{lcapgdmc:rules} 
\end{figure}

\newpage{}Figure \ref{lcapgdmc:rules}: Generation of a new molecule
using LCAP gradients for GDMC optimization. Here, only two sites are
considered. At each site, there are four possible fragments ($A1,A2,A3,A4$
for site $A$, and $B1,B2,B3,B4$ for site $B$). A gray box indicates
that the fragment is present in the current molecule. Based on the
LCAP gradients calculated for molecule $A1B1$, all of the fragments
at each site are reordered and the next molecule $A3B2$ is generated.
If $A3B2$ has been visited before, one site is randomly chosen to
undergo mutation. In this example, site $A$ is chosen and fragment
$A3$ is mutated to fragment $A1$ (the next highest ranking fragment)
and $A1B2$ is generated. This procedure is repeated until the maximum
number of iterations is reached or no more new molecules can be generated.

\begin{figure}[H]
\begin{centering}
\includegraphics{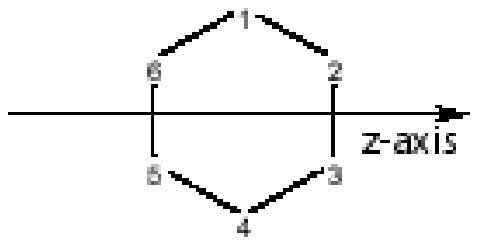} 
\par\end{centering}

\caption{}

\label{case1:frame} 
\end{figure}

Figure \ref{case1:frame}: Framework for Case I. Either a $CH$ or
$N$ can be placed at each site. 64 possible structures exist, without
considering symmetry equivalent structures.

\begin{figure}[H]
\begin{centering}
\includegraphics[width=15cm]{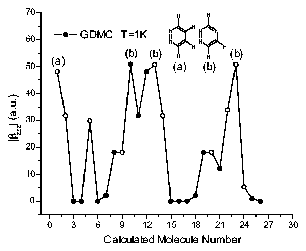} 
\par\end{centering}

\caption{}

\label{case1:opt} 
\end{figure}

Figure \ref{case1:opt}: Optimization profile for Case I (T=1K). The
x-axis indexes the calculated molecules during the optimization; the
y-axis represents the absolute value of $\beta_{zzz}$, where empty
circles denote negative molecular $\beta_{zzzz}$ values and solid
circles denote positive values. The initial molecule is structure
(a). Three degenerate molecules with structure (b) and the largest
$\beta_{zzz}$ value were obtained. During the optimization, the LCAP
gradients guided the generation of new molecules and resulted in a
greater likelihood of generating a new molecule with a higher property
value. The Monte Carlo random moves in the initial stage assisted
in jumping out of the local optimum. 

\begin{figure}[H]
\begin{centering}
\includegraphics{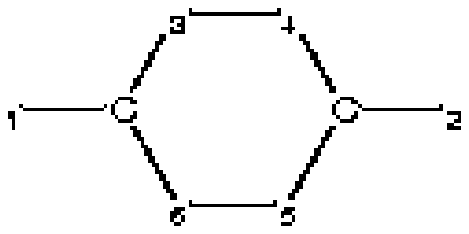} 
\par\end{centering}

\caption{}

\label{case2:frame} 
\end{figure}

Figure \ref{case2:frame}: Framework for Case II. At sites 1 and 2,
either an $H$, $CH_{3}$, $NO_{2}$ or $NH_{2}$ was placed. Either
a $CH$ or $N$ was placed at sites 3 to 6. Sites 1 and 2 are located
on the axis z. 256 possible structures exist, without considering
symmetry equivalent structures.

Figure \ref{case2:opt}: Optimization profiles for Case II with three
different temperature parameters. The largest $|\beta_{zzz}|$ molecule
(b) was found in all optimizations. When $T$ is increased from $1K$
to $50K$ to $100K$, the other chemically identical molecule (b)
was also found, suggesting that the optimization results do not depend
critically on the temperature parameter.

\begin{figure}[H]
\begin{centering}
\includegraphics[width=5cm]{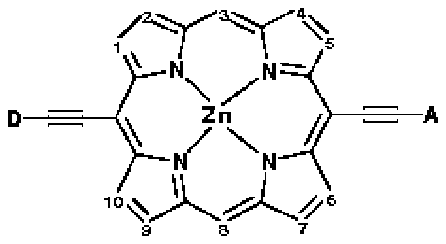} 
\par\end{centering}

\caption{}

\label{case3:frame} 
\end{figure}

Figure \ref{case3:frame}: Framework for Case III. At site $A$, one
of three acceptor groups ($NO_{2}$, $CN$ or $COH$) was placed.
At site $D$, one of three donor groups ($N(CH_{3})_{2}$, $CH_{3}$
or $OCH_{3}$) was placed. At sites 1 through 10, either a $CH$ or
$N$ was placed. 9,216 possible molecules exist, without considering
symmetry equivalent structures.

Figure \ref{case3}: (a) Initial molecule in Case III. (b) Molecule
with the largest $\beta_{zzz}$ value after four GDMC optimizations
were carried out using three different temperature parameters. (c)
Molecule with a normal porphyrin ring and the strongest donor and
acceptor groups in Case III.

\begin{figure}[H]
\begin{centering}
\includegraphics[width=15cm]{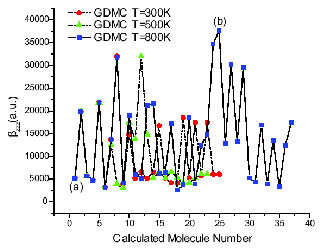} 
\par\end{centering}

\caption{}

\label{case3:opt} 
\end{figure}

Figure \ref{case3:opt}: Three optimization profiles for Case III
with three temperatures ranging from $300K$ to $800K$. $\beta_{zzz}$
values are always positive due to the donor-$\pi$-acceptor framework.
The rugged surface trapped three optimizations when $T=300K$ and
$500K$. At $T=800K$, GDMC jumped out of the local optimum and yielded
molecule (b), shown in Fig. \ref{case3}.

\newpage{}

\newpage{}

\newpage{}%

\newpage{}%

\newpage{}%

\newpage{}%

\newpage{}%

\newpage{}%

\end{document}